\documentclass[twoside]{scrartcl}
\usepackage[T1]{fontenc}
\usepackage[latin9]{inputenc}
\usepackage{units}
\usepackage{array}
\usepackage{graphicx}
\usepackage{epstopdf}
\usepackage{float}
\usepackage{amsmath}
\usepackage{amssymb}
\usepackage{esint}
\usepackage{multicol}
\usepackage{changepage}

\usepackage{xcolor}
\usepackage{float}
\usepackage{placeins}

\usepackage[justification=justified]{caption}
\usepackage{prettyref}
\newrefformat{sfig}{Supp. Fig. \ref{#1}}
\newrefformat{tab}{Supp. Table \ref{#1}}
\DeclareCaptionLabelFormat{supptable}{#1 S#2} 
\DeclareCaptionLabelFormat{suppfig}{#1 S#2} 

\usepackage[super,comma,sort]{natbib}

\usepackage{multibib}
\newcites{methods}{Online references}

\makeatletter

\pdfpageheight\paperheight
\pdfpagewidth\paperwidth

\makeatother

\begin{document}

\begin{flushleft}
{\Large
\textbf{Kalman inversion stress microscopy}
}
\newline
\\
V.~Nier\textsuperscript{1}, 
G.~Peyret\textsuperscript{2},
J.~d'Alessandro\textsuperscript{2},
S.~Ishihara\textsuperscript{4},
B.~Ladoux\textsuperscript{2,3} and
P.~Marcq\textsuperscript{1,*}
\\
\bigskip
\textbf{1} 
Laboratoire Physico Chimie Curie, Institut Curie, 
PSL Research University, Sorbonne Universit\'e, CNRS, 26 rue d'Ulm, 
75005 Paris, France\\
\textbf{2} Institut Jacques Monod, CNRS,  Universit\'e Paris Diderot, 75013 Paris, France\\
\textbf{3} Mechanobiology Institute, National University of Singapore, 
Singapore\\
\textbf{4} Graduate School of Arts and Sciences, The University of Tokyo, Tokyo 153-8902, Japan\\
\bigskip
* Corresponding author, philippe.marcq@curie.fr
\end{flushleft}

\bigskip
\centerline{August 29, 2018}
\date{}

\setcounter{section}{0}
\setcounter{figure}{0}
\setcounter{table}{0}
\setcounter{equation}{0}


\section*{Abstract}

Although mechanical cues are crucial to tissue morphogenesis and development,
the tissue mechanical stress field remains poorly characterized.
Given traction force timelapse movies, as obtained by traction force 
microscopy of \emph{in vitro} cellular sheets, we show that the tissue 
stress field can be estimated by Kalman filtering.
After validation using numerical data, we apply Kalman inversion stress 
microscopy  to experimental data. We combine the inferred stress
field with velocity and cell shape measurements to quantify the rheology 
of epithelial cell monolayers in physiological conditions, found to be close
to that of an elastic and active material.

\bigskip
\section*{Introduction}

The last decades have led to a growing awareness of the importance of 
mechanotransduction in cell and developmental biology 
\cite{Heisenberg2013,Iskratsch2014}. Noteworthy examples include the cell fate 
determination of stem cells as a function of their microenvironment 
\cite{Vining2017}, or the force-sensing machinery present
at adherens junctions \cite{Ladoux2015}.
Despite much recent effort \cite{Sugimura2016}, the measurement 
of internal forces in tissues remains challenging, and is often
limited to relative force estimates, as is the case for   
tissue scale ablation \cite{Bonnet2012} or geometry-based
inference \cite{Ishihara2012}.
Traction force microscopy (TFM), a well established tool of mechanobiology, 
allows to estimate \emph{in vitro} the force field exerted by cells and 
cell assemblies on their environment \cite{Style2014,Schwarz2015}.
Since internal stresses and external forces are balanced, a seemingly natural 
way to obtain the stress field from the traction force field would be to 
invert the force balance equation. This equation is however not invertible,
since three unknowns, the components of the symmetric stress tensor,
must be deduced from the two linear equations that correspond to the two 
components of the traction force. The problem may become invertible at the 
cost of postulating a tissue rheology \cite{Tambe2011}:
Monolayer Stress Microscopy (MSM) relies on the assumption of an elastic
tissue rheology, which is disputable given various evidence for   
viscous \cite{Guevorkian2010}, plastic \cite{Marmottant2009} 
and active \cite{Saw2017} behaviour  in tissues. Recently, we have 
described, validated  \cite{Nier2016}, and applied \cite{Saw2017}
Bayesian inversion stress microscopy (BISM), a 
stress inference method that dispenses with rheological hypotheses yet
allows to estimate the absolute value of the internal stress field of a cell 
sheet from an image of traction force measurement. In a Bayesian 
framework, the inferred stress is the mode of the posterior probability 
distribution function (pdf), given a prior stress distribution function 
equivalent to a regularizing term that controls the norm of the stress. 

Since TFM yields time-lapse movies that allowed to follow the time evolution 
of the traction force field, we adapt  Kalman filtering \cite{Kalman1960}
to this inversion problem \cite{Kaipio2006}, and formulate Kalman
inversion stress microscopy (KISM). The Kalman filter is, 
in a mean square error sense, an optimal estimator for 
Gaussian statistical models and remains the best linear estimator 
in the non-Gaussian case \cite{Anderson1979}.
Qualitatively, correlations between successive data frames are preserved
for high enough time resolution. Kalman filtering exploits this 
feature to obtain an estimate of the state variable while solving an 
under-determined inversion problem. 
Since KISM is free from any assumption concerning tissue mechanics,
we use the inferred stress field to study epithelial rheology.

\section*{Materials and methods}

\subsection*{Statistical model}
\label{sec:Kalman:stat}

Neglecting inertia, the force balance equation of a planar cell sheet exchanging
momentum with its substrate reads:
\begin{equation}
  \label{eq:forcebalance:hydro}
  \mathrm{div} \, \sigma = \vec{f}\,,
\end{equation}
where $\sigma(\vec{r},t)$ and $\vec{f}(\vec{r},t)$ respectively denote
the two-dimensional, symmetric stress tensor and traction force fields
at position $\vec{r}$ and time $t$.
On a discrete cartesian grid with spatial resolution $l$, 
we denote  $\nabla$ the discretized divergence (matrix) operator,
computed with fourth-order centered finite differences. 
Using the acquisition time as the time unit, $t$ becomes a discrete variable,
$t = 1,\ldots, T$, where $T$ is the total number of frames.
At each time step $t$, Eq.~(\ref{eq:forcebalance:hydro}) translates into
a system of coupled linear equations, where the unknown quantities are the
stress field components $\sigma_{xx}(i,j,t)$, $\sigma_{yy}(i,j,t)$ and
$\sigma_{xy}(i,j,t)$, with $i = 1,\ldots, N_x$, $j = 1,\ldots, N_y$,
using on a grid of size $N_x \, N_y$. 
For each point of coordinates $(i,j)$ on the spatial grid, solving this
system amounts to determining three unknowns given two equations,
one for each component of the traction force field 
$f_{x}(i,j,t)$ and $f_{y}(i,j,t)$: the system is not invertible, and calls 
for non-algebraic methods for its solution.

We perform Kalman inversion \cite{Kaipio2006}, and accordingly probabibilize 
the problem. Let $\vec{F}^t$ be the traction force vector encompassing both
traction force components on the whole spatial domain at time $t$:
\begin{equation}
  \label{eq:def:F}
  \vec{F}^t  =  [ f_x(1,1,t) \cdots f_x(N_x,N_y,t) \, 
f_y(1,1,t) \cdots f_y(N_x,N_y,t) ]^T \,,
\end{equation}
where the superscript $^T$ denotes the transpose. 
Let $\vec{\sigma}^t$ be a similar stress \emph{vector} 
including the three stress components at all grid points at time $t$:
\begin{equation}
  \label{eq:def:sigma}
  \vec{\sigma}^t  =  [ \sigma_{xx}(1,1,t) \ldots \sigma_{xx}(N_x,N_y,t) \, 
\sigma_{yy}(1,1,t) \ldots \sigma_{yy}(N_x,N_y,t) \,
\sigma_{xy}(1,1,t) \ldots \sigma_{xy}(N_x,N_y,t) ]^T \,.
\end{equation}
The observation model reads:  
\begin{equation}
  \label{eq:observation}
 \nabla  \vec{\sigma}^t = \vec{F}^t + \vec{\phi}^t
\end{equation}
with an additive, zero-mean, Gaussian observation noise 
$\vec{\phi}^t \sim \mathcal{N}(\vec{0}, s^2I)$ of variance $s^2$,
with correlations 
$\langle \phi_{\alpha}^t \phi_{\beta}^{t'} \rangle = 
s^2 \, \delta_{\alpha \beta} \, \delta_{t t'}$
for all components $\alpha,\beta = 1,\ldots,2 N_x N_y$.
The simplest possible expression of an evolution model for the stress
field is a random walk:
\begin{equation}
\label{eq:evolution:simple}
\vec{\sigma}^t= I \vec{\sigma}^{t-1}+ \vec{\xi}^t 
\end{equation}
where $I$ is the identity matrix, and the evolution noise $\vec{\xi}^t$ 
is zero-mean, Gaussian with variance $\gamma^2$,  
$\vec{\xi}^t \sim \mathcal{N}(\vec{0}, \gamma^2I)$, and correlations 
$\langle \xi_{\alpha}^t \xi_{\beta}^{t'} \rangle = 
\gamma^2 \, \delta_{\alpha \beta}\, \delta_{t t'}$,
$\forall \alpha,\beta = 1,\ldots,3 N_x N_y$.
In practice we implement the relevant stress-free boundary conditions in the 
evolution model, introducing an evolution matrix $B$:
\begin{equation}
\label{eq:evolution:full}
\vec{\sigma}^t= B \, \vec{\sigma}^{t-1}+ \vec{\xi}^t \,.
\end{equation}
The matrix $B$ is equal to the identity matrix $I$, except for diagonal
components set to zero due to stress-free boundary conditions.
In a confined system, we thus set $ \sigma_{ij} \, n_j=0$  at the boundary, 
up to the addition of the evolution noise, where $\vec{n}$ denotes the 
vector normal to the edge, and summation over repeated indices is implied.

At time $t=1$, the stress vector and its covariance matrix are respectively
initialized as $\hat{\vec{\sigma}}^1$ and $S^1$, as defined below. 
At times $t>1$, we combine 
traction force data $\vec{F}^t$ with the previous stress estimate 
$\hat{\vec{\sigma}}^{t-1}$ to compute iteratively $\hat{\vec{\sigma}}^t$, 
the filtered stress at time $t$,
with the matrix operations \cite{Anderson1979}:
\begin{eqnarray}
\label{eq:Kalman:matrix}
K^t&=& \left(B S^{t-1} B^T+\gamma^2 I \right) \nabla^T 
\left( \nabla \left(B S^{t-1} B^T+\gamma^2 I \right) \nabla^T+s^2I \right)^{-1}\\
\label{eq:Kalman:covariance}
S^t&=& \left(I-K^t\nabla\right) \left(B S^{t-1} B^T+\gamma^2 I \right)
\left(I-K^t\nabla\right)^T + s^2
K^t (K^t)^T \\
\label{eq:Kalman:sigma}
\hat{\vec{\sigma}}^t&=& B \hat{\vec{\sigma}}^{t-1}+K^t \left(\vec{F}^t-
\nabla B \hat{\vec{\sigma}}^{t-1}\right)
\end{eqnarray}
where $K^t$ is the Kalman matrix at time $t$.
The covariance matrix $S^t$ allows to determine error bars on the stress
estimate $\hat{\vec{\sigma}}^t$.

In physical terms, Eqs.~\eqref{eq:observation} and \eqref{eq:evolution:full} 
may be interpreted as follows \cite{Kaipio2006}. Let us denote 
$\mathcal{F}^t = \{  \vec{F}^1, \ldots,  \vec{F}^t\}$ the set of all 
traction force data up to time $t$. The goal of Kalman filtering
is to determine the conditional pdf $P( \vec{\sigma}^t | \mathcal{F}^t)$.
In the case of Gaussian statistics, this amounts 
to calculating its mean $\hat{\vec{\sigma}}^t$ and its covariance matrix  $S^t$.
One iteration step can be decomposed into a prediction step, followed
by an update step. The prediction step uses a Chapman-Kolmogorov equation 
to compute $P( \vec{\sigma}^t | \mathcal{F}^{t-1})$ given
$P( \vec{\sigma}^{t-1} | \mathcal{F}^{t-1})$ and a (Gaussian) Markov
transition kernel $P( \vec{\sigma}^t | \vec{\sigma}^{t-1})$,
determined by the evolution equation \eqref{eq:evolution:full}:
\begin{equation}
  \label{eq:prediction}
  P( \vec{\sigma}^t | \mathcal{F}^{t-1}) = \int \mathrm{d} \vec{\sigma}^{t-1}
\, P( \vec{\sigma}^t | \vec{\sigma}^{t-1})
\, P( \vec{\sigma}^{t-1} | \mathcal{F}^{t-1})  \,.
\end{equation}
The update step next deduces $P( \vec{\sigma}^t | \mathcal{F}^t)$
from Bayes formula: 
\begin{equation}
  \label{eq:update}
 P( \vec{\sigma}^t | \mathcal{F}^t) = \frac{P( \vec{F}^t | \vec{\sigma}^{t}) \, P( \vec{\sigma}^t | \mathcal{F}^{t-1})}{P( \vec{F}^t |\mathcal{F}^{t-1})} \,,
\end{equation}
where  $P( \vec{\sigma}^t | \mathcal{F}^{t-1})$ plays the role of a prior,
$P( \vec{F}^t | \vec{\sigma}^{t})$ is the likelihood determined by 
the observation equation \eqref{eq:observation}, and the 
denominator is a normalization factor
$P( \vec{F}^t |\mathcal{F}^{t-1}) = \int \mathrm{d} \vec{\sigma}^t
 P( \vec{F}^t | \vec{\sigma}^{t})  P( \vec{\sigma}^t | \mathcal{F}^{t-1})$.
In the Gaussian case, Eqs.~(\ref{eq:prediction}-\ref{eq:update}) lead to  
the iteration rule (\ref{eq:Kalman:matrix}-\ref{eq:Kalman:sigma}) 
\cite{Kaipio2006}. For an intuitive derivation of 
Eqs.~(\ref{eq:Kalman:matrix}-\ref{eq:Kalman:covariance}) in a simple,
scalar case, we refer the reader to \cite{Faragher2012}.

Concerning the algorithm's parameters, the observation noise variance $s^2$ is given by 
$s^2_{\mathrm{exp}}$, the experimental uncertainty on the traction 
force data. The evolution noise variance $\gamma^2$ is evaluated from data as
follows. Applying $\nabla$ to (\ref{eq:evolution:simple}) and substituting 
$\nabla \sigma^t$ using (\ref{eq:observation}), we obtain the evolution 
equation of the traction force: $\vec{F}^t=\vec{F}^{t-1}+\vec{\psi}^t$
where $\vec{\psi}^t = \nabla \vec{\xi}^t + \phi^{t-1} - \phi^t$, 
the sum of zero-mean Gaussian 
noises, is also a zero-mean Gaussian noise. Using the statistical independence 
in space and time between the different noises, we estimate the order of 
magnitude of the evolution noise variance:
\begin{equation}
  \label{eq:measure:gamma2} 
\gamma^2 \approx l^2  \,
\langle\langle(\vec{F}^t-\vec{F}^{t-1})^2\rangle\rangle + 4 l^2 s^2 
\end{equation}
where $\langle\langle...\rangle\rangle$ denotes spatial and temporal 
averaging. When inferring the stress field from experimental data, 
we typically use \cite{Nier2016} 
$s^2 = s_{\mathrm{exp}}^2 \approx 100 \, \mathrm{Pa}^2$ and find 
$\gamma^2 \approx 2 \, 10^5 \, \mathrm{Pa}^2 \mu\mathrm{m}^2$ (HaCaT cells),
$\gamma^2 \approx 3 \, 10^4 \, \mathrm{Pa}^2 \mu\mathrm{m}^2$  (MDCK cells)
when $l = \lambda = 25 \, \mu$m. 
For simplicity, we use as initial conditions 
$\hat{\vec{\sigma}}_{\alpha}^1 = \gamma$, $\forall \alpha$, 
and $S^1 = \gamma^2 I$.

\subsection*{Measures of accuracy}
\label{sec:Kalman:num:valid}

At each time step, KISM computes a set $\{\sigma^{\mathrm{inf}}\}$ of inferred 
stresses from the set of experimental traction forces $\{f^{\mathrm{exp}}\}$.
We calculate the inferred traction force field $\{f^{\mathrm{inf}}\}$ 
by applying the divergence operator to the inferred stress field:
$\vec{F}^{\mathrm{inf}} = \nabla \vec{\sigma}^{\mathrm{inf}}$. 
For each component $f_i = f_x, f_y$ of $\vec{f}$,
we calculate the coefficient of determination:
\begin{equation}
  \label{eq:def:R2:F}
  R_{i}^2(t) = 
1-\frac{\sum (f_{i}^{\mathrm{exp}}(t)-f_{i}^{\mathrm{inf}}(t))^2}
{\sum (f_{i}^{\mathrm{exp}}(t)- \langle f_{i}^{\mathrm{exp}}(t) \rangle)^2}\,.
\end{equation}
where the sums and the averages $\langle \ldots \rangle$ 
are performed over space. An aggregate quantifier $R_{f}^2$ of the accuracy of 
inference  for a given traction force movie is obtained by 
averaging $R_{i}^2(t)$ over time and over the components $x$ and $y$,
with the most accurate estimate corresponding to numerical 
values of $R_{f}^2$ closest to $1$.
The same measure of accuracy can be defined for numerical data,
replacing $\{f^{\mathrm{exp}}\}$ by $\{f^{\mathrm{num}}\}$ in the above expression.

Exact, spatially-averaged values of stress components can be 
calculated directly from traction force data in confined domains
where the boundary condition $\sigma_{ij} \, n_j=0$ applies.
Denoting as above spatial averages by brackets $\langle \ldots \rangle$,
we have $\langle \sigma_{ij}  \rangle =  - \langle f_i \, x_j \rangle$
\cite{landau1975elasticity}, 
with Cartesian coordinates $(x_1, x_2) = (x,y)$
(see also \cite{Nier2016} for an explicit derivation). 
Using this relation, we checked that the average inferred stress values  
$\langle \sigma_{ij}^{\mathrm{inf}}  \rangle$ 
agree with the values $\langle \sigma_{ij}^{\mathrm{exp}}  \rangle$
computed directly from the traction force data.

\subsection*{Numerical simulation}
\label{sec:Kalman:num}

For definiteness, we use the simulated traction force field of a 
compressible viscous tissue, obeying the constitutive equation:
\begin{equation}
\label{eq:stress:viscous}
\sigma=\eta \left(\vec{\nabla}\vec{v}+\left(\vec{\nabla}\vec{v} \right)^t \right)+\eta'\left(\vec{\nabla}.\vec{v} \right)I \,,
\end{equation}
with shear and bulk viscosities $\eta$ and $\eta'$,
interacting with its substrate through an effective fluid friction force
(friction coefficient $\xi_{d}$),
and driven by $n_d$ moving, active force dipoles:
\begin{equation}
\label{eq:traction}
\mathrm{div} \, \sigma =  \vec{f} = \xi_{d} \, \vec{v} - 
\sum_{n=1}^{n_d}\nabla. p^n(\vec{x},t)  \,.
\end{equation}
The dipole amplitudes increase towards the boundaries:
they are set proportional to $1+r/l_p$, where $r$ is the distance to the center
of the domain and $l_p = 5 \, \mathrm{\mu m}$ is a penetration length  
\cite{Roure2005}. 

The dynamics stems from the actively moving dipoles. 
Following \cite{Sepulveda2013}, we stipulate that the force dipoles 
tend to align their direction with their velocity, with the following 
relaxation equation of the orientation $\theta_d^n$ of the dipole towards 
the orientation $\theta_v^n$ of its velocity $\vec{v}^{\,n}$, 
\begin{equation}
\label{eq:dipole:angle}
\frac{d\theta_d^n}{dt}=- \frac{1}{\tau_d} (\theta_d^n-\theta_v^n)
\end{equation}
with a relaxation time $\tau_d$.
In addition, each dipole velocity $\vec{v}^n$ is given as an 
Ornstein-Uhlenbeck process 
\begin{equation}
\label{eq:dipole:velocity}
\frac{d\vec{v}^{\,n}}{dt}=-\frac{1}{\tau_d} \vec{v}^{\,n}+ \vec{\varphi}^{\,n}\\
\end{equation}
with correlation time $\tau_d$, and zero-mean, Gaussian white noise 
$\vec{\varphi}^{\,n}$ with correlations
$\langle \varphi^{\,n}_i(t) \varphi^{\,n}_j(t') \rangle =  s_v^2 \,
\delta_{ij} \delta(t-t')$ for components $i,j$.
Eqs.~(\ref{eq:dipole:angle}-\ref{eq:dipole:velocity}) determine 
the trajectories of dipoles, starting from random initial positions and 
orientations. The numerical resolution of 
Eqs.~(\ref{eq:stress:viscous}-\ref{eq:dipole:velocity})  
is performed with \texttt{FreeFem++} \cite{Hecht2012}.

We use material parameter values typical of epithelial cell monolayers 
\cite{Nier2016}$^,$\cite{Sepulveda2013,Roure2005}:
friction coefficient $\xi_{d} = 10^0 \, \mathrm{kPa \,\mu m^{-1}s}$, 
shear viscosity $\eta= 10^4 \, \mathrm{kPa\,\mu m\,s}$, 
bulk viscosity $\eta'=\eta$, $n_d = 100$ dipoles with
a typical amplitude $1$ kPa, 
a correlation time $\tau_d = 10^4$ s, 
a noise amplitude $s_v=7 \,10^{-4} \, \mathrm{\mu m \,s^{-3/2}}$.
The simulated tissue is confined in a square of area 
$100 \times 100 \,\mathrm{\mu m}^2$, with a spatial resolution of 
$l=2\,\mathrm{\mu m}$. We include movies of
this simulation for the traction force (in $\mathrm{kPa}$) and for the stress 
(in $\mathrm{kPa \,\mu m}$) over a total duration of $3\,\mathrm{h}$
and with a time step of $30\,\mathrm{s}$ (see Movies S1 and S2).

The numerical resolution of the set of equations given above
immediately yields a numerical data set of stresses 
$\{ \sigma^{\mathrm{num}}\}$. 
To account for the measurement error,
we add to the simulated traction force field a zero-mean, Gaussian white 
noise of amplitude $s^{\%}_{\mathrm{exp}} \, f_{\mathrm{max}}$, where $f_{\mathrm{max}}$ is the maximal value of the norm of the traction force,
and obtain a numerical data set $\{f^{\mathrm{num}}\}$ of traction forces.
The set $\{\sigma^{\mathrm{inf}}\}$ of inferred stresses is next computed
from $\{f^{\mathrm{num}}\}$ with KISM. 
At time $t$, we 
calculate for each component of the stress $\sigma_{ij} = \sigma_{xx},
\sigma_{yy}, \sigma_{xy}$ the 
coefficient of determination
\begin{equation}
  \label{eq:def:R2:sigma}
  R_{ij}^2(t) = 
1-\frac{\sum (\sigma_{ij}^{\mathrm{num}}(t)-\sigma_{ij}^{\mathrm{inf}}(t))^2}
{\sum (\sigma_{ij}^{\mathrm{num}}(t)- \langle \sigma_{ij}^{\mathrm{num}}(t) \rangle)^2}
\end{equation}
An aggregate quantifier $R^2_{\sigma}$ of the accuracy of inference 
is obtained by averaging first over the stress components $xx$, $yy$ and $xy$ at
time $t$ (coefficient of determination $R^2_{\sigma}(t)$), and then over time.  
As above, the evolution noise variance is computed from averaged traction force
increments (Eq.~\eqref{eq:measure:gamma2}), with typical values
$\gamma^2 \approx 10^{5} \, \mathrm{Pa}^2 \mu\mathrm{m}^2$
and we use as initial conditions $\hat{\vec{\sigma}}_{\alpha}^1 = \gamma$, 
$\forall \alpha$, and $S^1 = \gamma^2 I$.

The Kalman inversion that yielded stresses in Movie S2 from 
the traction force data in Movie S1 required $150$ min of Intel-Xeon E5 CPU. 
As the main computational bottleneck of the algorithm are
matrix inversions in \eqref{eq:Kalman:matrix}, we expect the computational 
cost to scale as $O(N^3)$ \cite{Murphy2012}.

\subsection*{Experimental methods}
\label{sec:Kalman:experiment}

Cells were cultured in Dulbecco's Modified Eagle Medium, 
supplemented with $10\%$ of fetal veal serum and $1\%$ of 
penicillin-streptomycin at $37^{\circ}$C, with $5\%$ $\mathrm{CO_2}$.
For experiments, cells were concentrated at around 4 million cells per mL 
and a drop of 200 $\mathrm{\mu L}$ was added in the medium of the experimental 
Petri dish. Incubation time lasted from $15$ to $30$ min depending on the 
concentration needed at the beginning of the experiment and on the cell line. 
Cells that did not attach were then washed and the substrate incubated 
overnight. Cells were confined in a $500 \times 500 \, \mu \mathrm{m}^2$ 
square domain. The  height of the monolayer was of the order of 
$h_{\mathrm{HaCaT}} \simeq 3 \, \mu$m and 
$h_{\mathrm{MDCK}} \simeq 5 \, \mu$m\cite{Nier2016} for HaCaT and MDCK cells, 
respectively.

To confine the cells to square patterns, we used micro-contact 
printing on soft gel, as previously described \cite{Vedula2014}. 
Briefly, PDMS stamps exhibiting square features were incubated with 
a fibronectin solution at $75\, \mathrm{mg \, mL}^{-1}$ for $45$ min. 
After rinsing with water, the dried stamp was put in contact with the 
surface of a polyvinyl-alcohol membrane (Sigma Aldrich). Then the 
membrane was put upside-down on the soft gel to allow the transfer 
of the protein from the membrane to the gel. Finally, the membrane 
was dissolved with a $2\,\%$ Pluronics-F27 (Sigma Aldrich) solution 
which was let to incubate for $2$ h to make the non-printed areas 
anti-adhesive for the cells.

Live imaging was performed with a $10$X objective on a BioStation
IM-Q (Nikon) at $37^{\circ}$C and $5\%$ CO$_2$ with humidification. 
Images of the cells were taken every $10$ min. The velocity fields were 
calculated by Particle Image Velocimetry (PIV) analysis with MATPIV 1.6.1, 
a Matlab (the Mathworks) implemented script. An interrogation window of 
$64$ pixels (approximately $41 \,\mu$m) was selected with an overlap of $75\,\%$. 
Vectors higher than a speed threshold manually determined were removed, 
and a local median filter was applied.

Moving epithelial cells exerted traction forces on their substrate that 
can be calculated from the displacement field of the substrate. 
This was achieved using $200$ nm fluorescent beads attached on the surface 
as previously described \cite{Saw2017,Peyret2016}. A Z-stack of images was 
taken every $10$ min. Images were first processed with ImageJ to obtain the best 
focus plane for each time point (Stack Focuser plugin), then stabilized
(Image Stabilizer plugin), and background beads were removed. The substrate 
displacements were measured with PIV, using interrogation areas of 
$15.5\times 15.5\,\mathrm{\mu m^2}$  with an overlap of $75\%$. 
Images of fluorescent beads were acquired and compared with an 
image of the gel at rest, obtained at the end of the experiment. 
A drop of $200\,\mathrm{\mu L}$ of sodium dodecyl sulfate 
($0.1 \,\mathrm{g \, mL}^{-1}$) was added in order to lyse and detach the cells. 
Forces were calculated from the displacement field by 
Fourier Transform Traction Cytometry (FTTC) with an 
open-source ImageJ plugin developed by Qingzong Tseng \cite{Martiel2015}. 
The gel had a 
Young modulus of $15$ kPa, with a Poisson ratio of $0.5$. We used 
a regularization parameter of $10^{-10}$ for FTTC.
The gel thickness ($\approx 200\,\mu$m) was large compared
to the typical correlation lengths of both the beads' displacements 
($\approx 85\,\mu$m) and the traction forces ($\approx 14\,\mu$m), 
so that finite thickness corrections to the TFM could be safely neglected.

To obtain the cell shape tensor $Q$ in a cell sheet, we followed the same 
procedure as described in \cite{Saw2017}. A clear image was obtained, with 
individual cell boundaries visible. The image was smoothed using Bandpass 
Filter  in ImageJ to remove unnecessary details. The filter size of small 
structures was set to roughly one-third the size of a single cell. 
The ImageJ plugin OrientationJ was used to detect the direction of the 
largest eigenvector of the structure tensor of the image \cite{Rezakhaniha2012} 
for each pixel (for a window size of roughly one-quarter the size of a
single cell).
The output is the orientation angle $\theta_Q$ with values 
ranging from $-90^{\circ}$ to+ $+90^{\circ}$. 
The local cell shape tensor tensor, $Q$, was calculated  for each point 
on a grid that discretized the image, using an in-house Matlab code,
averaging over pixel directions in a fixed-size region that contained 
3-5 cells: 
\begin{equation}
  \label{eq:def:Q}
Q=\left\langle
\begin{pmatrix}
\cos^2\theta_Q - 1/2& \cos\theta_Q \sin\theta_Q\\
\cos\theta_Q \sin\theta_Q& \sin^2\theta_Q - 1/2
\end{pmatrix}
\right\rangle,
\end{equation}
where the brackets denote averaging over a $64$\,pixels-wide window, 
with $75\%$ overlap.
Only pixels that resided in the region of the cell body were taken into 
account for this calculation (white regions obtained by
Auto Local Threshold function in ImageJ) as cell
boundary regions could have orientations that are perpendicular 
to the cell body.

\subsection*{Data analysis}
\label{sec:Kalman:data}

All fields were interpolated on the TFM grid
using Matlab's interp2 function, and coarse-grained over boxes of
linear extension $\lambda = 25 \, \mu$m, excluding a domain of width $\lambda$
along each boundary.
We checked that results of the rheological analysis did not change
for a larger coarse-graining scale $\lambda = 50 \, \mu$m.
Since the solution first needs to
relax to its optimum (Figs.~\ref{fig:numerical_validation},
\ref{fig:data:stress:HaCaT} and \ref{fig:data:stress:MDCK}), 
stress estimates at short time may be unreliable. For this reason,
we conservatively discarded the first $10$ h of the estimated stress
when estimating rheological parameters. 

Tensors were decomposed into the sum of a deviatoric (traceless)
term and of an isotropic term, as in 
$\sigma = \mathrm{dev}\, \sigma + \frac{1}{2} \mathrm{tr} \, \sigma \, I$,
where tr denoted the trace (\emph{i.e.} 
$\mathrm{tr} \, \sigma = \sigma_{xx} + \sigma_{yy}$).
The symmetrized velocity gradient tensor $D$ was defined as:
 \begin{equation}
\label{eq:def:D}
 D = 
 \begin{pmatrix}
 \partial_x v_x & \frac{1}{2} \left( \partial_x v_y + \partial_y v_x \right) \\
 \frac{1}{2} \left( \partial_x v_y + \partial_y v_x \right) & \partial_y v_y
 \end{pmatrix}\,.
 \end{equation}
Spatial derivatives were computed with Matlab's gradient function
on the scale $\lambda$. 
Time derivatives denoted with a dot were total derivatives, as in:
\begin{equation}
  \label{eq:def:dot}
  \dot{\sigma}_{xx} = \left( \frac{\partial}{\partial t}  + 
\vec{v} . \vec{\nabla} \right) \sigma_{xx} \,.
\end{equation}

To quantify plithotactic behavior, we estimated the angles $\theta_{\sigma v}$ 
between the direction of the velocity field and the principal axis 
of the stress tensor $\sigma$; 
and $\theta_{\sigma Q}$ between the principal axes of $\sigma$ and $Q$.
Following Ref.~\cite{Tambe2011}, we computed angular distributions
corresponding to the highest quintile of the stress anisotropy.
Relaxing this thresholding condition did not change our
results qualitatively, but increased the dispersion
and made angular distributions wider.
Angular distributions were fitted by a zero-mean von Mises distribution
of parameter $\kappa \ge 0$:
\begin{equation}
  \label{eq:def:vonMises}
  p_{\mathrm{von \, Mises}}(\theta) = \frac{e^{\kappa \, \cos 2 \theta}}{2 \pi \, I_0(\kappa)}
\end{equation}
where $\theta$ belonged to the interval $[-\frac{\pi}{2} \, \frac{\pi}{2}]$ and 
$I_0$ denoted the modified Bessel function of order zero.
A smaller value of $\kappa$ corresponds to a wider 
distribution \cite{Fisher1995}, which is uniform when $\kappa = 0$.

\section*{Results}

The state and observation variables of the 
Kalman filter were defined as the two-dimen\-sio\-nal stress and traction force 
fields $\sigma(\vec{r},t)$ (in kPa $\mu$m)  and $\vec{f}(\vec{r},t)$ 
(in kPa), where $\vec{r}$ and $t$ denoted position and time 
(Materials and methods, Statistical model). 
The observation equation (Eq.~(\ref{eq:observation}))
was the linear, two-dimensional
force balance equation, $\mathrm{div} \, \sigma = \vec{f}$, 
discretized on a grid of finite mesh, and supplemented with an additive
observation noise, assumed to be Gaussian and white.
With applications to confined cellular sheets in mind, the stress-free
boundary conditions were implemented in the evolution equation, where the 
stress fields at two consecutive time steps differed by an additive evolution
noise, also Gaussian and white (Eq.~(\ref{eq:evolution:full})).
The stress estimate was iteratively updated as an optimal combination of  
its estimate at the previous time step with the contribution of traction 
force data at the current time step
(Eqs.~(\ref{eq:Kalman:matrix}-\ref{eq:Kalman:sigma})). 
We emphasize that, contrary to Bayesian inversion, Kalman inversion did 
not require a prior.

\begin{figure}[!t]
\centering 
\begin{adjustwidth}{-0.5in}{} 
\includegraphics[scale=0.45]{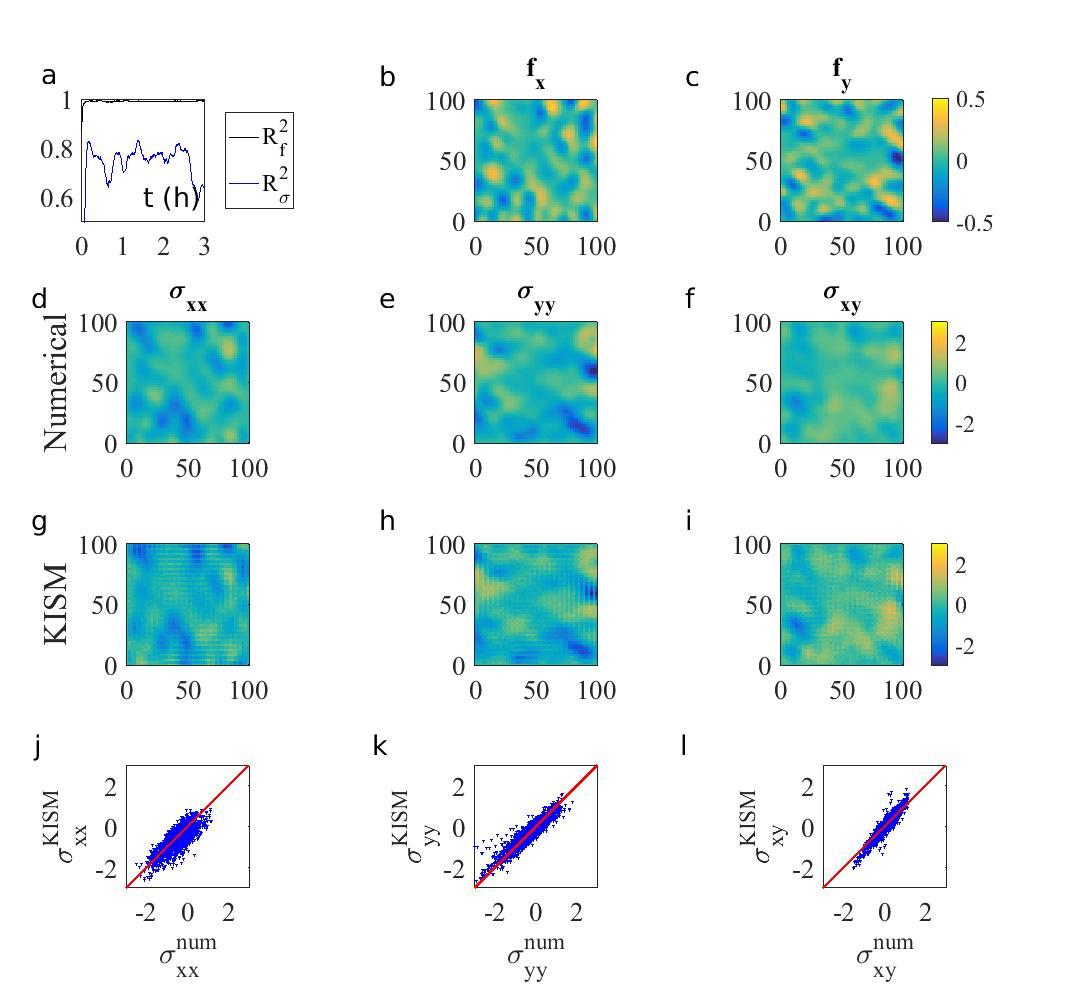}
\end{adjustwidth}
\caption{\textbf{Numerical validation.}
(a) $R^2_f$ and $R^2_{\sigma}$ \emph{vs.} time $t$.
(b-c) Heat maps at $t = 1\, \mathrm{h}$ of the components $f_x$ and $f_y$  of 
the simulated traction force field $\vec{f}^{\mathrm{num}}$  (unit: kPa).
Panels d-i: Heat maps at $t = 1$ h of the components $\sigma_{xx}$, $\sigma_{yy}$ 
and $\sigma_{xy}$ of: 
(d-f) the simulated stress field $\sigma^{\mathrm{num}}$;
(g-i) the stress field $\sigma^{\mathrm{KISM}}$ inferred with KISM  
 (unit: $\mathrm{kPa\,\mu m}$).
Panels j-l: Comparison between inferred  KISM stress $\sigma^{\mathrm{KISM}}$ and
true simulated stress $\sigma^{\mathrm{num}}$. The red line is the bisector 
$y = x$. The relative noise amplitude is $s^{\%}_{\mathrm{exp}}=10\%$.
Time unit: h; length unit: $\mu$m.
\label{fig:numerical_validation} 
}
\end{figure}

\begin{figure}[!t]
\centering 
\begin{adjustwidth}{-0.8in}{-0.65in} 
\includegraphics[scale=0.4]{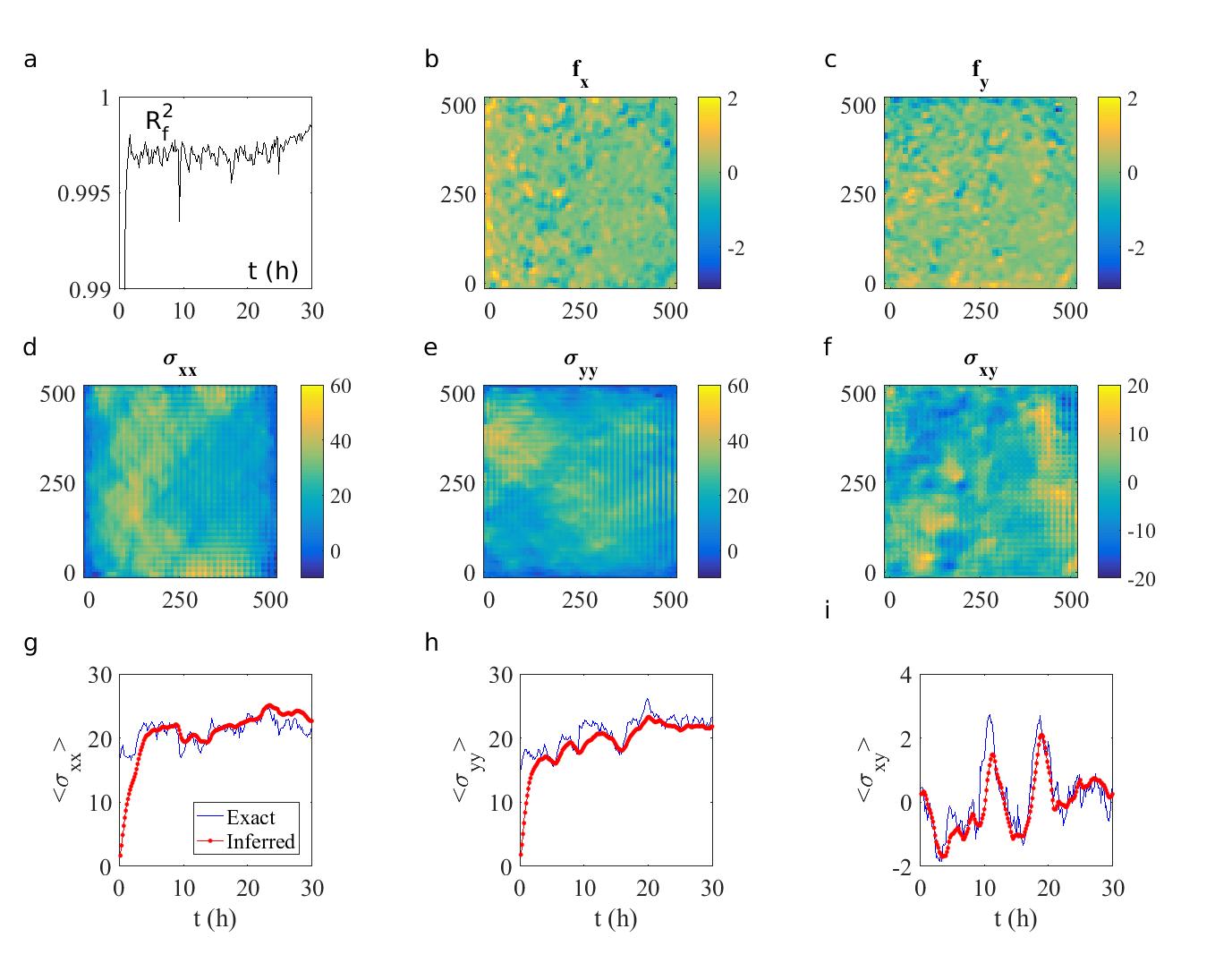}  
\end{adjustwidth}
\caption{\textbf{Epithelial stress field: HaCaT cells.}
(a) Coefficient of determination $R^2_f$ \emph{vs.} time $t$.
(b-c) Heat maps of the components $f_{x}$ and $f_{y}$  of the
experimental traction force field  at $t = 17$ h (unit: kPa).
(d-f) Heat maps of the components $\sigma_{xx}$, $\sigma_{yy}$ and 
$\sigma_{xy}$ of the stress field inferred with KISM   
at $t = 17$ h (unit: $\mathrm{kPa\,\mu m}$). 
 (g-i) Spatially-averaged stress components \emph{vs.} time $t$.
Blue line: exact values computed from the first moment of the traction force
field, $\langle \sigma_{ij}  \rangle =  - \langle f_i \, x_j \rangle$. 
Red dots: values estimated by KISM.
Globally  ($N = 8$), the average inferred pressure was negative, 
$\langle\langle \pi\rangle\rangle^{\mathrm{HaCaT}} = - 23.9 \pm 2.4$ kPa $\mu$m
between $t = 10$h and $t = 30$h, while the average inferred shear stress was
$\langle\langle \sigma_{xy}\rangle\rangle^{\mathrm{HaCaT}} = 0.2 \pm 0.8$ 
kPa $\mu$m in the presence of sustained oscillations \cite{Peyret2016}.
Time unit: h; length unit: $\mu$m.
\label{fig:data:stress:HaCaT} 
}
\end{figure}

\begin{figure}[!t]
\centering 
\begin{adjustwidth}{-0.5in}{} 
\includegraphics[scale=0.4]{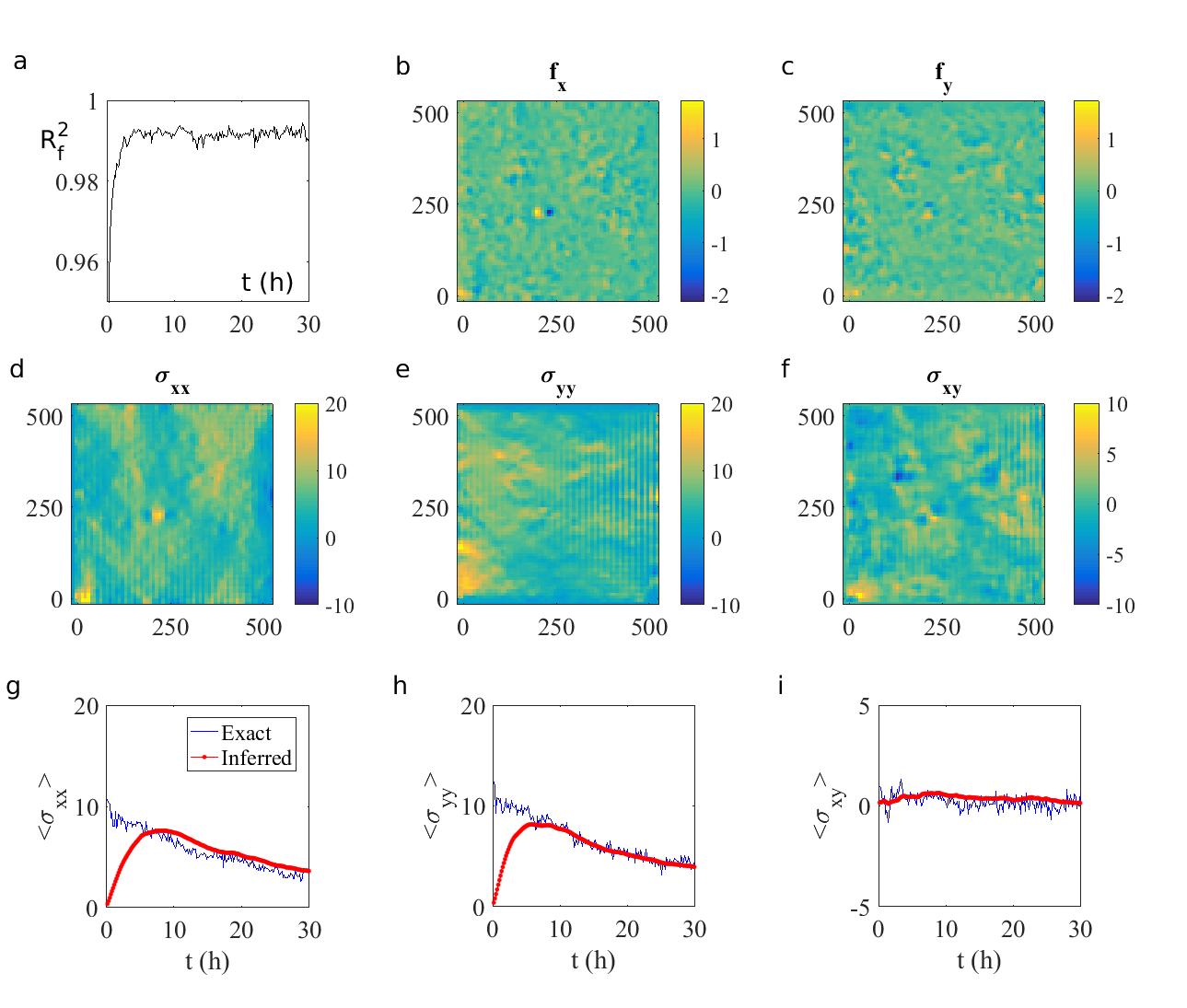}  
\end{adjustwidth}
\caption{\textbf{Epithelial stress field: MDCK cells.}
(a) Coefficient of determination $R^2_f$ \emph{vs.} time $t$.
(b-c) Heat maps of the components $f_{x}$ and $f_{y}$  of the
experimental traction force field  at $t = 17$ h (unit: kPa).
(d-f) Heat maps of the components $\sigma_{xx}$, $\sigma_{yy}$ and 
$\sigma_{xy}$ of the stress field $\sigma^{\mathrm{inf}}$ inferred with KISM   
at $t = 17$ h (unit: $\mathrm{kPa\,\mu m}$).
(g-i) Spatially-averaged stress components  \emph{vs.} time $t$.
Blue line: exact values computed from the first moment of the traction force
field, $\langle \sigma_{ij}  \rangle =  - \langle f_i \, x_j \rangle$.  
Red dots: values estimated by KISM. Globally ($N = 5$), the average 
inferred pressure was negative between $t = 10$ h and $t = 30$ h, 
$\langle\langle\pi\rangle\rangle^{\mathrm{MDCK}} = -6.3 \pm 0.6$ kPa $\mu$m,
while the average inferred shear stress was consistent with $0$:
$\langle\langle\sigma_{xy}\rangle\rangle^{\mathrm{MDCK}} = 0.2 \pm 0.2$ 
kPa $\mu$m. Time unit: h; length unit: $\mu$m.
\label{fig:data:stress:MDCK} 
}
\end{figure}

\FloatBarrier

We first validated KISM using traction force and stress data obtained
by the numerical resolution of a simple model of a cellular sheet as a 
compressible, viscous material driven by active, motile force dipoles
(Materials and methods, Numerical simulations and Movies~S1-2). 
As shown in Fig.~\ref{fig:numerical_validation} when the relative noise
amplitude was $s^{\%}_{\mathrm{exp}} = 10 \, \%$, our statistical model
allowed to estimate accurately the simulated stress, with coefficients of 
determination $R^2_f \simeq 0.99$ and $R^2_{\sigma} \simeq 0.75$. 
We checked that the accuracy of inference was insensitive to parameter 
values of the model, such as the correlation time $\tau_d$ and noise amplitude 
$s_v$  that control the time evolution of force dipoles.
After a relaxation regime, typically shorter than ten frames,
the dynamical rule (\ref{eq:Kalman:matrix}-\ref{eq:Kalman:sigma}) 
converged towards an accurate estimate of the stress field 
(Fig.~\ref{fig:numerical_validation}). 
As expected from the 
asymptotic stability properties of Kalman filters \cite{Anderson1979},
the dynamics converged rapidly
towards the same asymptotic state when we modified the initial condition as
$\hat{\vec{\sigma}}^1 = c \, \gamma$ and $S^1 = c^2 \,\gamma^2 I$,
$c = 10^{-3}$ and $c = 10^{3}$ (Figs.~S\ref{fig:numerical_robustness}ab).
The L2-norm $\Arrowvert K^t \Arrowvert$ of the Kalman gain matrix also 
converged rapidly towards its asymptotic value in the same conditions,
although on a slightly slower time scale 
(Fig.~S\ref{fig:numerical_robustness}b).
Varying the values of the noise variances $\gamma^2$,
we observed that  $R^2_{\sigma}$ exhibited a shallow optimum close
to the estimate $\gamma^2 \approx 10^{-1} \, \mathrm{Pa}^2 \mu\mathrm{m}^2$
(Fig.~S\ref{fig:numerical_robustness}c).
Finally, we verified that the accuracy of inference was a
decreasing function of $s^{\%}_{\mathrm{exp}}$ 
(Fig.~S\ref{fig:numerical_robustness}d).
Unsurprisingly, we observed that the accuracy of inference is a decreasing 
function of the time resolution (Fig.~S\ref{fig:numerical_robustness}e), 
as well as of the scale over which the data may be coarse-grained
(Fig.~S\ref{fig:numerical_robustness}f).

From experimental traction force timelapse movies, we inferred the 
stress field of monolayers of human keratinocytes (HaCaT cells,
Fig.~\ref{fig:data:stress:HaCaT}, Movies~S3-S5) and of 
Madin-Darby canine kidney  cells (MDCK, Fig.~\ref{fig:data:stress:MDCK},
Movies S7-S9), see Materials and methods, Experimental methods.
We considered the first $30$ hours after confluence, during which 
sustained collective motion was observed (Movies~S3 and S7)
and ignored the jammed state arising later due to an increase in cell 
density. The accuracy of stress inference was quantified by the coefficient 
of determination $R^2_f$ that compares the experimental traction force data 
with  an ``inferred'' traction force field computed as the divergence of 
the inferred stress field. The value of $R^2_f$ was always larger than 
$0.98$ after a brief relaxation regime
(Figs.~\ref{fig:data:stress:HaCaT}a and \ref{fig:data:stress:MDCK}a).
Comparing the spatially-averaged stress values to their expected values
(Materials and methods, Measures of accuracy), we also confirmed that the 
absolute value of the stress field was correctly estimated
(Figs.~\ref{fig:data:stress:HaCaT}g-i and \ref{fig:data:stress:MDCK}g-i).
Following Kalman filtering, the average inferred values 
are smoother than the true behavior. However, the stress computed 
at early time points may be highly inaccurate. We note that the relaxation 
time towards an accurate inference (a few hours) is similar 
for $R^2_{\mathrm{f}}$ and $\langle \sigma_{ij}\rangle$.
As expected, both cell sheets were under tension (negative average pressure
$\pi = -(\sigma_{xx} + \sigma_{yy})/2$).

\begin{figure}[!t]
\centering 
\includegraphics[scale=0.25]{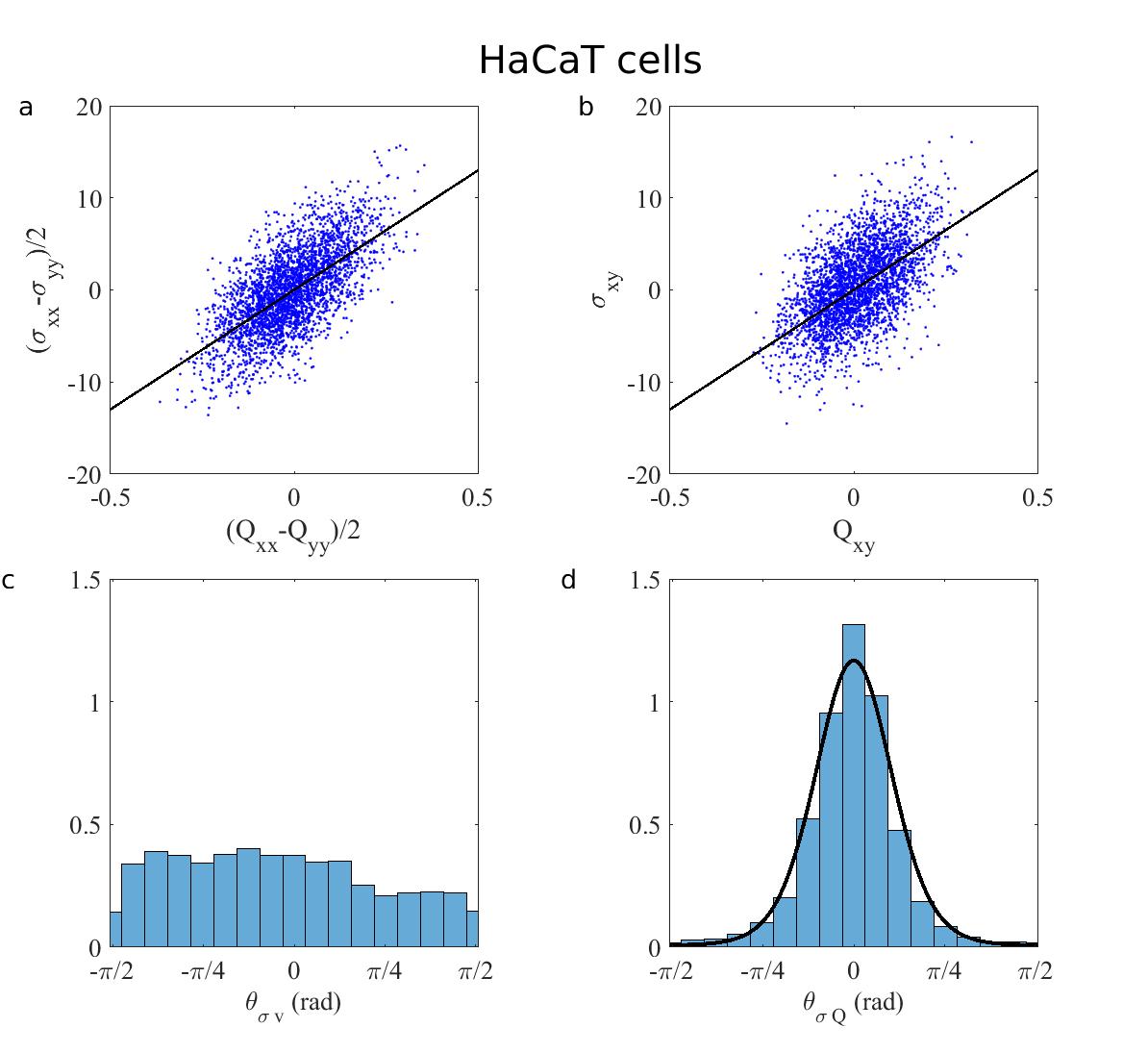} 

\vspace*{-0.25cm}
\includegraphics[scale=0.25]{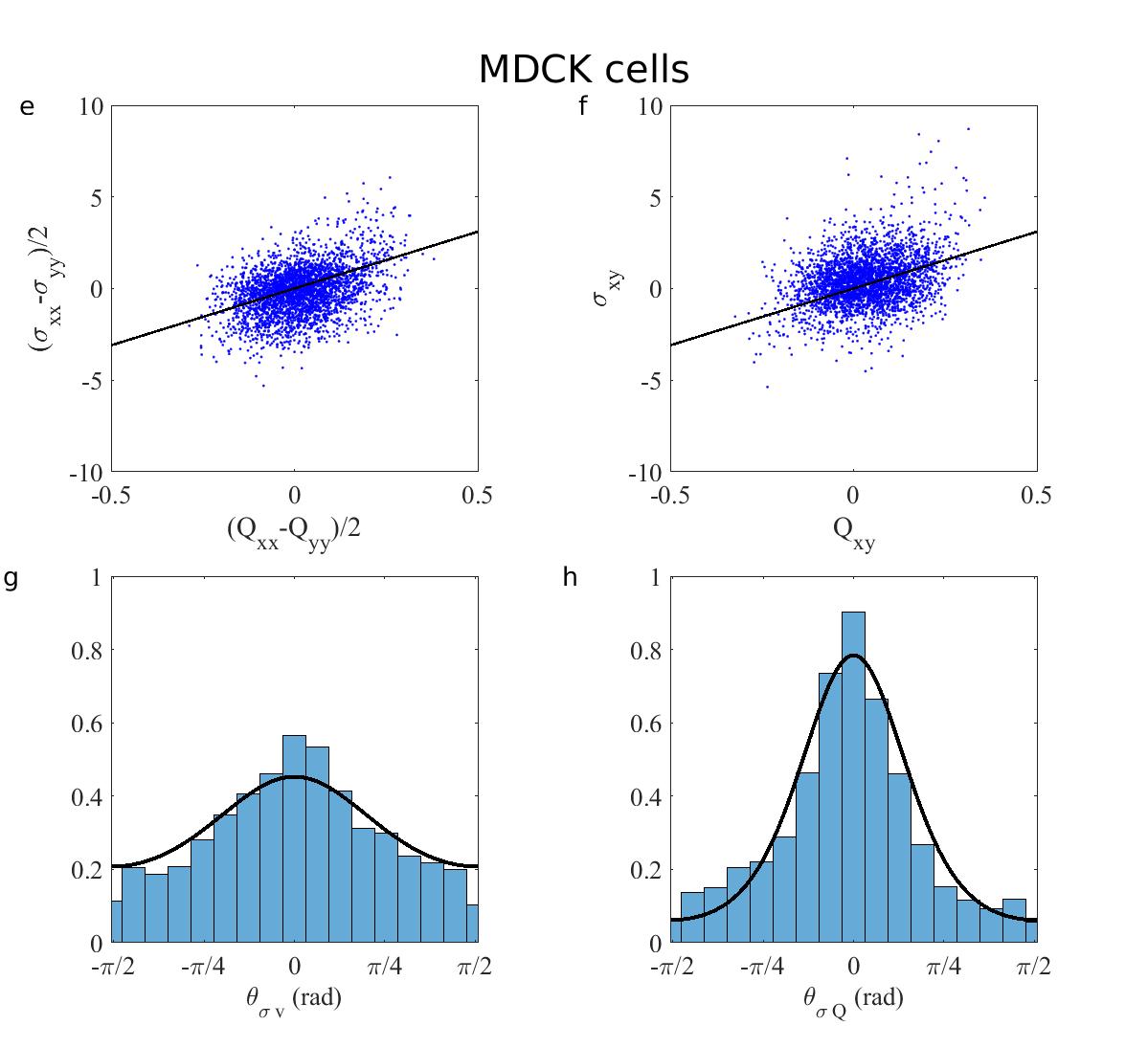} 
\caption{\textbf{Epithelial rheology. (a-d) HaCaT cells.}
(a-b) Components of the deviatoric stress tensor (unit: $\mathrm{kPa\,\mu m}$)
\emph{vs.} same components of the cell shape tensor. The slope of the
black lines is the average parameter value $\zeta = 26.0\,\mathrm{kPa\,\mu m}$
($N=8$). (c-d) Pdfs of the angles $\theta_{\sigma v}$ and $\theta_{\sigma Q}$,
given a stress anisotropy belonging to its highest quintile.
The black curve is a  zero-mean von Mises distribution
with the average parameter value $\kappa_{\sigma Q} = 2.46$ ($N=8$).
\textbf{(e-h) MDCK cells.}
Same representations as for HaCaT cells, with parameter values
$\zeta = 6.2\,\mathrm{kPa\,\mu m}$, $\kappa_{\sigma v} = 0.39$,
$\kappa_{\sigma Q} = 1.27 $ ($N = 5$).
\label{fig:data:rheo} 
}
\end{figure}

In order to characterize mechanical behavior at the scale of the tissue, 
all fields were coarse-grained over a mesoscopic scale $\lambda = 25 \, \mu$m. 
Motivated by previous work on epithelial rheology \cite{Saw2017,Ishihara2017}, 
we focused on the cell shape tensor $Q$ and the symmetrized velocity gradient 
tensor $D$ (Materials and methods, Data analysis). Surprisingly,
neither the stress tensor nor its time derivative were significantly
correlated with $D$ (Tables~1,2). 
Since the measured cell-shape (nematic) tensor $Q$ was traceless, we focused on
the deviatoric stress tensor, dev $\sigma$, which exhibited strong positive
correlations with $Q$ (Figs.~\ref{fig:data:rheo}ab-ef).
A linear regression of our data with the constitutive equation:
$  \mathrm{dev} \, \sigma =  \zeta \; Q$
allowed to measure the  material parameters 
$\zeta^{\mathrm{HaCaT}} = 26.0 \pm 0.3 \, \mathrm{kPa} \, \mu\mathrm{m}$ 
(N = 8) and $\zeta^{\mathrm{MDCK}} = 6.2 \pm 0.7 \, \mathrm{kPa} \, \mu\mathrm{m}$
(N = 5). Importantly, linear regressions for the two components of the 
deviatoric tensors yielded consistent slopes, in agreement with 
tensor symmetry. Elastic stress in epithelia is expected to be 
proportional to the cell shape tensor \cite{Ishihara2017}. In addition, 
symmetries allow an active contribution to the same  relationship, with an 
active parameter $\zeta_{\textrm{a}}$.
We may thus define an effective shear elastic modulus $G$ from
the relation $\zeta = G - \zeta_{\textrm{a}}$, with $\zeta_{\textrm{a}} >0$
for extensile active materials such as an MDCK monolayer \cite{Saw2017}.
Although estimating $G$ is here impractical, we note that the
  order of magnitude found for $\zeta$ is compatible with estimates of elastic
  moduli derived from 
the force-extension curve of suspended cell monolayers \cite{Harris2012}.
In our data, the presence of dissipative behavior was suggested by 
correlations between dev $\sigma$ and the time derivatives  dev $\dot \sigma$  
and $\dot Q$ (Table~1), where the dot denotes a total derivative.
However, these correlations were typically smaller than $0.15$, 
confirming that deviatoric stress depended dominantly on cell shape.
Rheological behavior of the epithelial cell sheets was, to first order, 
that of an active and elastic material.

Finally, the measured stress field allowed to characterize plithotactic 
behaviour, defined as the tendency of cells to align their velocity
with the principal axis of the tissue stress tensor during collective cell 
migration \cite{Tambe2011}. To quantify this tendency,
we measured the angle $\theta_{\sigma v}$ between the tissue velocity
and the principal axis of the stress tensor. For MDCK cells, its pdf could 
be fitted by a zero-mean von Mises distribution of parameter 
$\kappa^{\mathrm{MDCK}}_{\sigma v} = 0.39 \pm 0.12$ 
(Fig.~\ref{fig:data:rheo}g).
However, HaCaT cells did not exhibit plithotactic behaviour as the distribution
of $\theta_{\sigma v}$ was nearly uniform (Fig.~\ref{fig:data:rheo}c). 
Plithotactic behaviour, which may be related to cell-cell junction 
and cytoskeleton remodelling, was cell-type dependent\cite{Tambe2011}.
For both cell types, the distribution 
of the angle $\theta_{\sigma Q}$ between the principal axes of the stress 
and cell shape tensors was strongly peaked close to $0$, and could be fitted
by a zero-mean von Mises distribution 
with parameters $\kappa^{\mathrm{HaCaT}}_{\sigma Q} =  2.46 \pm 0.43$
and $\kappa^{\mathrm{MDCK}}_{\sigma Q} =   1.27 \pm 0.16 $.
The cell shape tensor had an orientation close to that of the 
stress tensor (Figs.~\ref{fig:data:rheo}d \ref{fig:data:rheo}h,
Movies S6 and S10).

\begin{table*}[!t]
   \centering
\begin{adjustwidth}{-0.6in}{0in} 
\begin{tabular}{cccccccc}
  \hline 
      & $(\mathrm{dev} \, \sigma, Q)$ & $(\mathrm{dev} \sigma, \mathrm{dev} D)$ 
& $(\mathrm{dev} \, \sigma, \mathrm{dev} \, \dot \sigma)$ 
& $(\mathrm{dev} \,  \sigma, \dot Q)$ 
& $(\mathrm{dev} \, \dot \sigma, \mathrm{dev} \, D)$ \\
  \hline
    HaCaT  & $\quad 0.59 \pm 0.06 \quad$ & $\quad 0.01 \pm 0.09 \quad$
& $\quad -0.01 \pm 0.02 \quad$ 
& $\quad -0.11 \pm 0.04 \quad$ 
& $\quad -0.01 \pm 0.07 \quad$  \\
  \hline
    MDCK  & $\quad 0.43 \pm 0.03 \quad$ & $\quad -0.02 \pm 0.04 \quad$
& $\quad -0.07 \pm 0.03 \quad$ 
& $\quad -0.13 \pm 0.02 \quad$ 
& $\quad 0.01 \pm 0.02 \quad$  \\
  \hline
\end{tabular}
\caption{
\label{tab:robust:sim} 
\textbf{Correlation coefficients: deviators of tensors.}\\
HaCaT cells: $N=8$; MDCK cells: $N=5$.
}
\end{adjustwidth}
\end{table*}

\begin{table*}[!t]
   \centering
\begin{tabular}{cccccc}
  \hline
      & $(\mathrm{tr} \, \sigma, \mathrm{tr} \, D)$ 
& $(\mathrm{tr} \, \sigma, \mathrm{tr} \, \dot \sigma)$  
& $(\mathrm{tr} \, \dot \sigma, \mathrm{tr} \, D)$ \\
  \hline
    HaCaT  & $\quad 0.01 \pm 0.09 \quad$ 
           & $\quad -0.01 \pm 0.03 \quad$ 
           & $\quad 0.01 \pm 0.09 \quad$  \\
  \hline
    MDCK  & $\quad -0.03 \pm 0.05 \quad$ 
          & $\quad -0.07 \pm 0.11 \quad$ 
          & $\quad -0.02 \pm 0.01 \quad$  \\
  \hline
\end{tabular}
\caption{
\label{tab:robust:sim} 
\textbf{Correlation coefficients: traces of tensors.}\\
HaCaT cells: $N=8$; MDCK cells: $N=5$.
}
\end{table*}

\FloatBarrier

\section*{Discussion}

To summarize, Kalman filtering led to accurate estimates of epithelial
stress,  without any assumption on tissue mechanical properties. 
We expect that its applicability and reliability would not differ
in the case of time-dependent \cite{Blanch-Mercader2017}, or 
spatially inhomogeneous \cite{Rodriguez-Franco2017} rheological properties.
The only assumption underlying KISM,
that the observation and evolution noises are Gaussian processes, 
could be relaxed using particle filtering, 
where arbitrary pdfs are sampled by Markov 
chain Monte Carlo methods. We opted for the simpler, Gaussian case given 
the accuracy of estimates thus obtained.
Note that a spatial prior may be used when Kalman filtering \cite{Kaipio2006}, 
closely following Bayesian inversion stress microscopy: in fact BISM may even 
be used to set the initial condition \cite{Nier2016}. A drawback is the need to
introduce a dimensionless regularisation parameter $\Lambda$, 
whose value must be determined from data \cite{Nier2016b}. 
We opt here for simplicity, and propose an inference method that does not 
require prior information, and therefore does not explicitly regularize 
the solution. Note that KISM estimates are generally in good agreement 
with BISM estimates (Fig.~S\ref{fig:data:stress:BISM}).
However KISM cannot be applied to single images, by construction,
and its accuracy decreases with the time resolution of the traction force movie.
Otherwise,
the conditions for applicability of KISM are the same as for
BISM, since the two inference methods share the same physical constraints.
Although the cell sheet should in principle be flat, and be characterized
by an approximately  constant height $h(\vec{r},t)$ for a description 
in terms of two-dimensional stress to be valid, this last assumption may 
be relaxed, as KISM may be implemented by replacing the traction force 
$f$ by the ratio $\vec{f}(\vec{r},t)/h(\vec{r},t)$
and inferring the three-dimensional stress $\sigma_{3D} = \sigma/h$   
from $\mathrm{div} \, \sigma_{3D} = \vec{f}/h$ as above.

In principle, Kalman smoothing \cite{Kaipio2006}$^,$\cite{Sarkka2013}
may further improve the accuracy of inference, as information contained
in frames posterior to that of the current estimate is also taken
into account, in addition to the forward evolution in time. 
We implemented a Kalman, fixed-interval smoother for stress inference, 
but found that the  improvement compared to the Kalman filter was marginal, 
except for the first time steps \cite{Nier2016b}, where the accuracy of 
Kalman filtering is limited. Since the additional computational cost required 
by smoothing is substantial, mostly in terms of memory allocation, we also 
opt for simplicity concerning this aspect, and favor Kalman filtering, 
rather than Kalman smoothing for stress estimation.  

A geometry-based Bayesian inversion method has been developed
to infer the stress field of flat epithelial cell sheets \emph{in vivo},
for instance in the \emph{Drosophila} pupa, using the positions of cell 
vertices and orientations of cell junctions as input data 
\cite{Ishihara2012}$^,$\cite{Ishihara2013}. 
We believe that Kalman filtering could also be applied to geometry-based 
inference so as to dispense with the need of a prior.

Remarkably, cell shape anisotropy was a good, zero-th order 
proxy for deviatoric stress, up to a cell type-dependent scale parameter
$\zeta$ that we measured.
This observation is consistent with the model proposed by some of us 
\cite{Ishihara2017} in the limit of an active and elastic rheology,
provided that cell rearrangements are rare. This result, obtained here for 
deviatoric stress in 2D, is reminiscent of that obtained for 1D stress 
in expanding MDCK monolayers \cite{Vincent2015}. 
Of note, the active viscous model shown in \cite{Blanch-Mercader2017} 
to explain quantitatively epithelial cell monolayer expansion 
can also be interpreted  as an elastic and active model
in this quasi-1D geometry.

The evolution equation \eqref{eq:evolution:full}
may be made more complex than a random walk to test the relevance 
of a given rheological model. Recent work \cite{Kondo2018} combined 
a Rauch-Tung-Striebel smoother with an Expectation-Maximization algorithm 
to infer the elastic moduli of an expanding cell monolayer, found in the 
kPa $\mu$m range. However, a natural extension of the approach
to Maxwell's model of a viscoelastic liquid led to overfitting and failed
to yield estimates of monolayer viscosities. Both results are consistent
with our observations.

Epithelial rheology could be quantified
in physiological conditions, where the ranges of forces 
and of deformations are not determined by an external operator,
but by the spontaneous activity of the cells.
The HaCaT cell sheet was likely stiffer than the MDCK cell sheet:
Determining the molecular cause of this difference is an open question
that we would like to address in the future. This work paves the 
way towards the inference of the constitutive equations of 
\emph{in vitro} cellular sheets from experimental data.

\section*{Acknowledgements}

Financial supports from the European Research Council
under the European Union's Seventh Framework Programme (FP7/2007-2013) / ERC
grant agreement 617233 (BL), the LABEX "Who am I?", and the Agence National de
la Recherche (POLCAM, ANR-17-CE13-0013) are gratefully acknowledged.


\newpage

\centerline{\Large
\textbf{Supporting Material}
}
\bigskip
\bigskip

\pagestyle{empty}


\section*{Movies}

\begin{itemize}
\item Movie S1. Numerical simulation: traction force, $x$ and $y$ components.
  Length unit: $\mu$m. Force unit: kPa.
  Time resolution: $\Delta t = 30$ s.
  Spatial resolution: $\Delta x = 2 \,\mu$m.
Duration: $3$ h.

\item Movie S2. Numerical simulation: stress, $xx$, $yy$ and $xy$ 
components.
Length unit: $\mu$m. Stress unit: kPa $\mu$m.
Time resolution: $\Delta t = 30$ s.
Spatial resolution: $\Delta x = 2 \,\mu$m.
Duration: $3$ h.
  
\item Movie S3. Experimental data: HaCaT cells.
Scale bar: $100 \, \mu$m.

\item Movie S4. Experimental data: HaCaT cells, traction force, 
$x$ and $y$ components.
Length unit: $\mu$m. Force unit: kPa.
Time resolution: $\Delta x = 10$ min.
Spatial resolution: $\Delta x = 3.9 \,\mu$m.
Duration: $20$ h.

\item Movie S5. Inferred data: HaCaT cells, stress, 
$xx$, $yy$ and $xy$ components.
Length unit: $\mu$m.
Stress unit: kPa $\mu$m.
Time resolution: $\Delta t = 10$ min.
Spatial resolution: $\Delta x = 3.9 \,\mu$m.
Duration: $20$ h.

\item Movie S6. Rheology:  HaCaT cells.
Left: velocity; center: principal axis of the deviatoric stress
tensor dev $\sigma$; right: principal axis of the cell shape tensor $Q$
(arbitrary units).
Time resolution: $\Delta t = 10$ min.
Spatial resolution: $\lambda = 25 \,\mu$m.
Duration: $20$ h.

\item Movie S7. Experimental data: MDCK cells.
Scale bar: $100 \, \mu$m.
  
\item Movie S8. Experimental data: MDCK cells, traction force,
$x$ and $y$ components.
Length unit: $\mu$m. Force unit: kPa.
Time resolution: $\Delta t = 10$ min.
Spatial resolution: $\Delta x = 3.9 \,\mu$m.
Duration: $20$ h.

\item Movie S9. Inferred data: MDCK cells, stress, 
$xx$, $yy$ and $xy$ components.
Length unit: $\mu$m. Stress unit: kPa $\mu$m.
Time resolution: $\Delta t = 10$ min.
Spatial resolution: $\Delta x = 3.9 \,\mu$m.
Duration: $20$ h.

\item Movie S10. Rheology, MDCK cells.
Left: velocity; center: principal axis of the deviatoric stress
tensor dev $\sigma$; right: principal axis of the cell shape tensor $Q$
(arbitrary units).
Time resolution: $\Delta t = 10$ min.
Spatial resolution: $\lambda = 25 \,\mu$m. 
Duration: $20$ h.

\end{itemize}

\newpage
\FloatBarrier
\setcounter{figure}{0}
\setcounter{table}{0}

\begin{figure}[!t]
\centering 
  \includegraphics[scale=0.5]{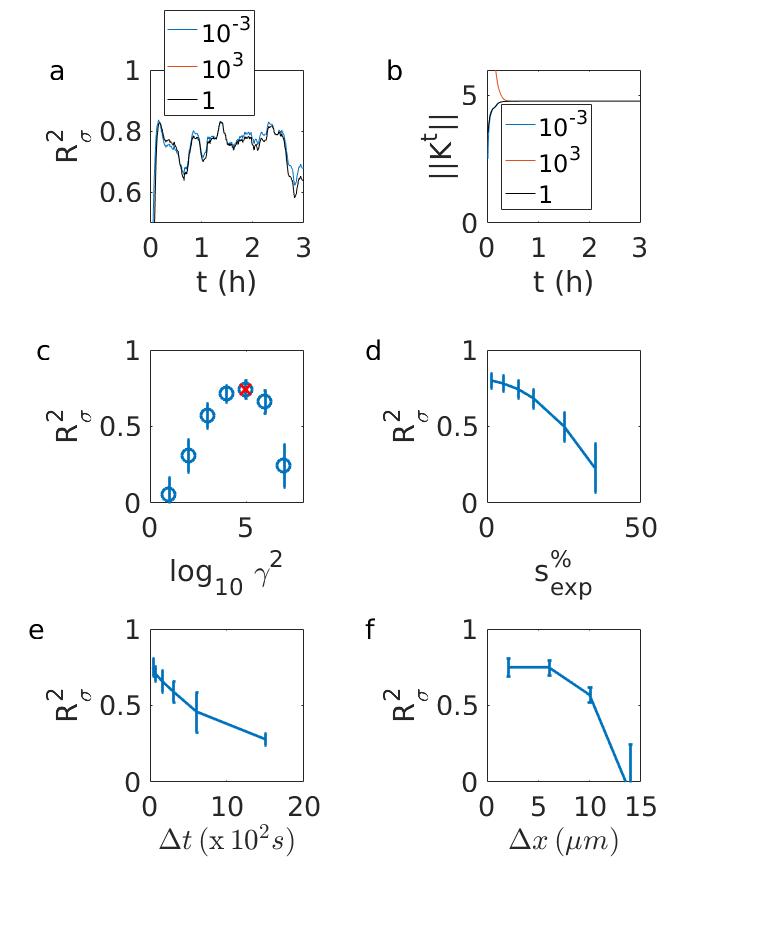}
\captionsetup{labelformat=suppfig}
  \caption{\textbf{Robustness.} (a-b)
$R^2_{\sigma}$ and $||K^t||$ \emph{vs.} time $t$ (unit: h)
for initial conditions differing by a multiplicative factor
$c = 10^{-3}, 1, 10^{3}$. The case $c=1$ corresponds
to the inference performed as in Fig.~\ref{fig:numerical_validation},
and is used in other panels of this figure. 
(c) $R^2_{\sigma}$ \emph{vs.} $\gamma^2$.
The maximum of $R^2_{\sigma}$ is close to the estimated value 
$\gamma^2 = 9.2 \, 10^4 \, \mathrm{Pa}^2 \, \mu \mathrm{m}^2$ (red cross) 
set as described in the Materials and methods, and used in other panels of 
this figure. (d) $R^2_{\sigma}$ \emph{vs.} $s^{\%}_{\mathrm{exp}}$.
The noise level in other panels is $s^{\%}_{\mathrm{exp}} = 10\,\%$.
(e) $R^2_{\sigma}$ \emph{vs.} time resolution $\Delta t$.
The time resolution in other panels is $\Delta t = 30$ s.
(f) $R^2_{\sigma}$ \emph{vs.} spatial resolution $\Delta x$.
The spatial resolution in other panels is $\Delta x = 2 \,\mu$m.
\label{fig:numerical_robustness} 
}
\end{figure}

\begin{figure}[!t]
\centering 
\includegraphics[scale=0.6]{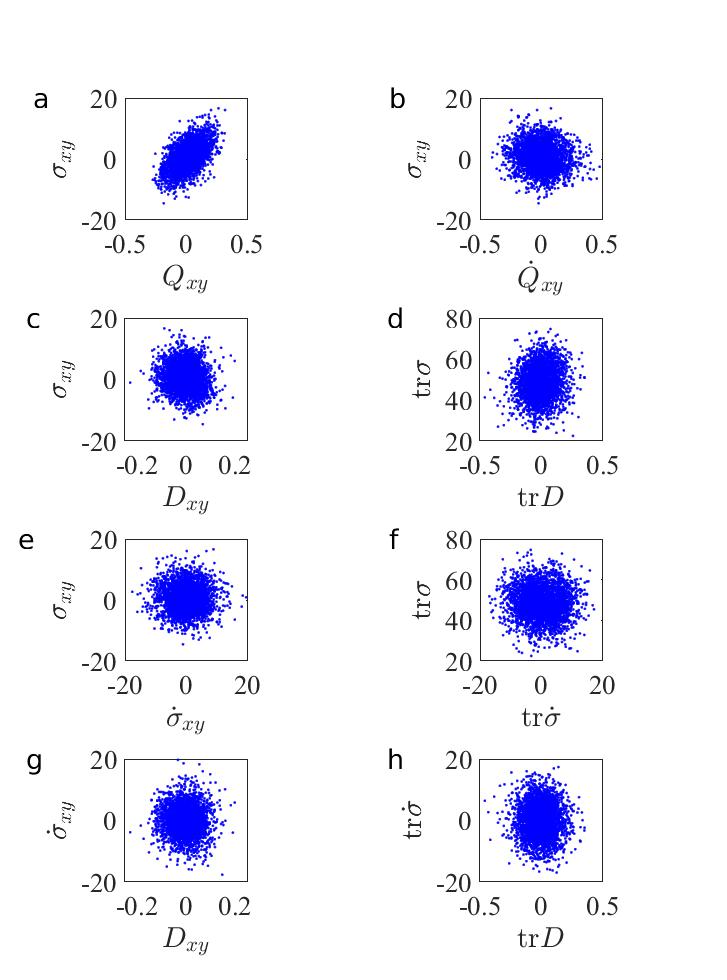} 
\captionsetup{labelformat=suppfig}
\caption{
\textbf{Scatter plots: HaCaT cells. Shear component and trace.}
\textbf{a} $\sigma_{xy}$ \emph{vs.} $Q_{xy}$;
\textbf{b} $\sigma_{xy}$ \emph{vs.} $\dot{Q}_{xy}$;
\textbf{c} $\sigma_{xy}$ \emph{vs.} $D_{xy}$;
\textbf{d} $\mathrm{tr} \, \sigma$ \emph{vs.} $\mathrm{tr} \, D$;
\textbf{e} $\sigma_{xy}$ \emph{vs.} $\dot{\sigma}_{xy}$;
\textbf{f} $\mathrm{tr} \, \sigma$ \emph{vs.} $\mathrm{tr} \, \dot{\sigma}$;
\textbf{g} $\dot{\sigma}_{xy}$ \emph{vs.} $D_{xy}$;
\textbf{h} $\mathrm{tr} \, \dot{\sigma}$ \emph{vs.} $\mathrm{tr} \, D$;
Stress unit: $\mathrm{kPa\,\mu m}$. Time unit: h.
Velocity gradient unit: h$^{-1}$.
Same data as in Fig.~\ref{fig:data:rheo}. The average correlation
coefficients are given in Tables~1 and 2 ($N=8$).
\label{fig:data:corr:HaCaT} 
}
\end{figure}

\begin{figure}[!t]
\centering 
\includegraphics[scale=0.6]{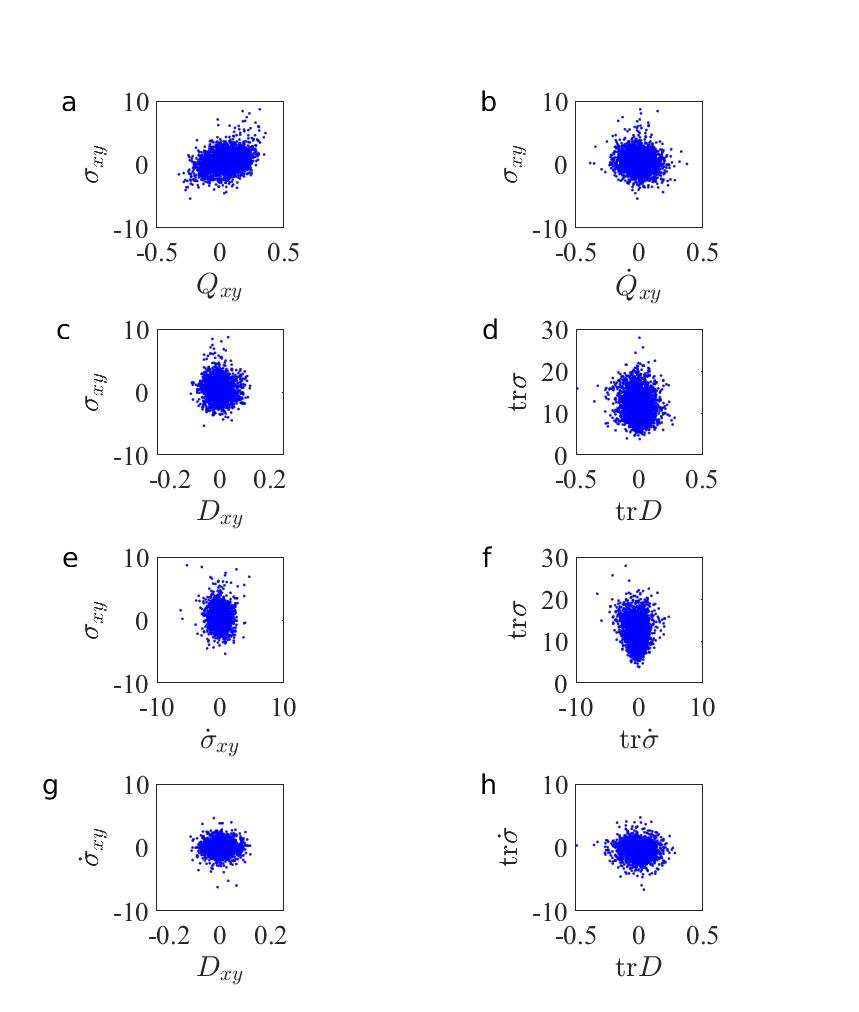} 
\captionsetup{labelformat=suppfig}
\caption{
\textbf{Scatter plots: MDCK cells. Shear component and trace.}
\textbf{a} $\sigma_{xy}$ \emph{vs.} $Q_{xy}$;
\textbf{b} $\sigma_{xy}$ \emph{vs.} $\dot{Q}_{xy}$;
\textbf{c} $\sigma_{xy}$ \emph{vs.} $D_{xy}$;
\textbf{d} $\mathrm{tr} \, \sigma$ \emph{vs.} $\mathrm{tr} \, D$;
\textbf{e} $\sigma_{xy}$ \emph{vs.} $\dot{\sigma}_{xy}$;
\textbf{f} $\mathrm{tr} \, \sigma$ \emph{vs.} $\mathrm{tr} \, \dot{\sigma}$;
\textbf{g} $\dot{\sigma}_{xy}$ \emph{vs.} $D_{xy}$;
\textbf{h} $\mathrm{tr} \, \dot{\sigma}$ \emph{vs.} $\mathrm{tr} \, D$;
Stress unit: $\mathrm{kPa\,\mu m}$. Time unit: h.
Velocity gradient unit: h$^{-1}$.
Same data as in Fig.~\ref{fig:data:rheo}. The correlation coefficients
are  given in Tables~1 and 2 ($N=5$).
}
\label{fig:data:corr:MDCK} 
\end{figure}

\begin{figure}[!t]
\centering 
\begin{adjustwidth}{-1.0in}{-0.65in} 
\includegraphics[scale=0.4]{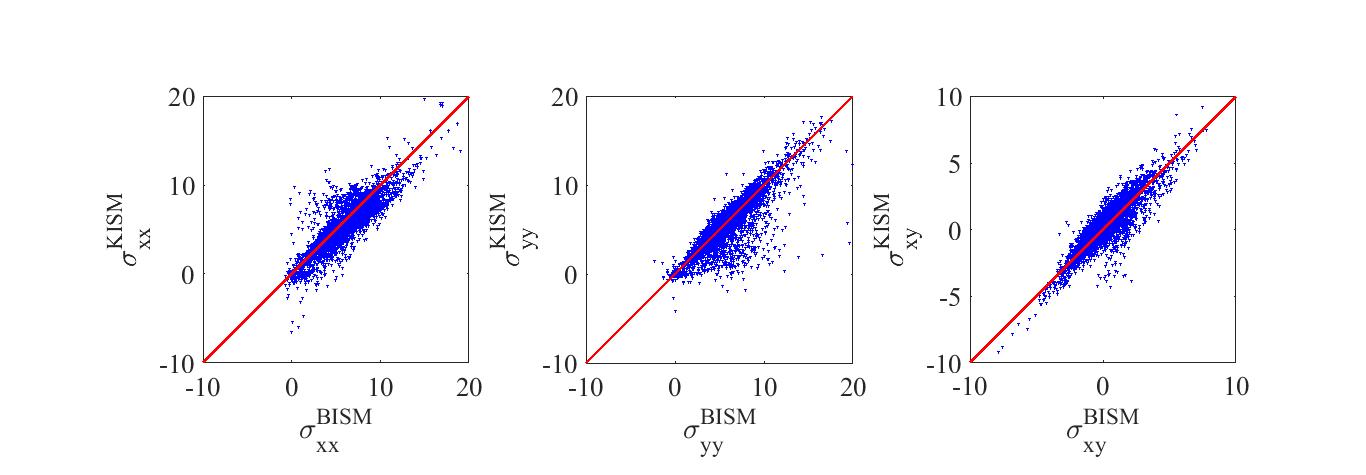} 
\end{adjustwidth}
\captionsetup{labelformat=suppfig}
\caption{\textbf{Comparison with BISM.}
Components of the KISM stress (as in Fig.~\ref{fig:data:stress:MDCK},
MDCK cells) \emph{vs.}  corresponding components of the 
BISM stress (regularisation parameter $\Lambda = 10^{-6}$).
The red lines are the bisectors $y = x$. Unit: $\mathrm{kPa\,\mu m}$.
\label{fig:data:stress:BISM} 
}
\end{figure}

\end{document}